\documentclass{emulateapj}
\usepackage{amsmath}
\usepackage{apjfonts}
\usepackage{natbib,graphicx}
\usepackage{longtable}
\newcommand{\cp}{\citep}
\newcommand{\ct}{\citet}

\begin{document}

\title{How Thermal Evolution and Mass Loss Sculpt Populations of Super-Earths and Sub-Neptunes: Application to the Kepler-11 System and Beyond}

\author{Eric D. Lopez}
\author{Jonathan J. Fortney$^1$}
\altaffiliation[$^1$]{Alfred P. Sloan Research Fellow}
\author{Neil Miller}
\affil{Department of Astronomy and Astrophysics, University of California, Santa Cruz, CA 95064} 

\begin{abstract}

We use models of thermal evolution and XUV-driven mass loss to explore the composition and history of low-mass low-density transiting planets. We investigate the Kepler-11 system in detail and provide estimates of both the current and past planetary compositions. We find that a H/He envelope on Kepler-11b is highly vulnerable to mass loss. By comparing to formation models, we show that in situ formation of the system is extremely difficult. Instead we propose that it is a water-rich system of sub-Neptunes that migrated from beyond the snow line. For the broader population of observed planets, we show that there is a threshold in bulk planet density and incident flux above which no low-mass transiting planets have been observed. We suggest that this threshold is due to the instability of H/He envelopes to XUV-driven mass loss.  Importantly, we find that this mass loss threshold is well reproduced by our thermal evolution/contraction models that incorporate a standard mass loss prescription.  Treating the planets' contraction history is essential because the planets have significantly larger radii during the early era of high XUV fluxes.  Over time low mass planets with H/He envelopes can be transformed into water-dominated worlds with steam envelopes or rocky super-Earths.  Finally, we use this threshold to provide likely minimum masses and radial velocity amplitudes for the general population of \textit{Kepler} candidates. Likewise, we use this threshold to provide constraints on the maximum radii of low-mass planets found by radial velocity surveys.

\end{abstract}

\subjectheadings{planetary systems; planets and satellites: composition, formation, interiors, physical evolution; stars: individual (Kepler-11)}

\section{Introduction}
In recent years, the frontier of the search for extrasolar planets has pushed towards ever smaller and more Earth-like worlds. We now know of dozens of Neptune mass planets and have even found the first definitively rocky extrasolar planets \cp{Batalha2011,Leger2009}. In between, transit searches have begun finding a population of low-mass low-density ``super-Earths''. Beginning with the discovery of GJ1214b \cp{Charbonneau2009}, these planets represent a new class of exoplanets that do not have any analog in our Solar System. Basic questions about their composition, structure, and formation are still unknown. Are these, in fact, scaled up versions of the Earth that simply have thick hydrogen/helium envelopes atop of rock/iron cores? Or are they instead scaled down versions of Neptune that are rich in water and other volatile ices?

The distinction between water-poor super-Earths or water-rich sub-Neptunes has fundamental implications for how these planets formed. So far these low-mass low-density (hereafter LMLD) planets have only been found well inside the snow-line. If these planets only contain rock, iron, and hydrogen/helium, then it is possible they formed close to their current orbits \cp{Hansen2011}. However, if a significant fraction of their mass is in water, then they must have formed beyond the snow-line and migrated in to their current locations \cp{Alibert2011,Ida2010,Rogers2011}.

The Kepler-11 system \cp{Lissauer2011a} is an extremely powerful tool for exploring the features of LMLD planets. With six transiting planets orbiting a close solar analog, it is the richest extrasolar system currently known. Moreover, five of the planets have masses from Transit Timing Variations (TTVs), and all five of these fall into the low-mass low-density regime in between Earth and Neptune. These five planets are all interior to Mercury's orbit, with periods from 10 to 47 days. This provides a unique laboratory to test the possible composition, formation, and evolution of LMLD planets and how these vary as a function of both period and planet mass.

Transiting planets with measured masses, like those in Kepler-11, are particularly valuable because we can determine their mean density. All the planets in Kepler-11 have densities too low for pure rock, and therefore must have some sort of thick envelope of volatiles. Likewise, all the planets except Kepler-11b are less dense than pure water and so must have at least some hydrogen/helium. 

Unfortunately, mass and radius alone cannot uniquely determine a planet's composition. In general, there is a large degeneracy between the relative amounts of rock, iron, water, and hydrogen/helium \cp{Rogers2010a}. This problem is particularly acute for planets with radii $\approx 2-4 \: R_{\mathrm{\oplus}}$, since in this range any of these four constituents can be important. Indeed these sorts of degeneracies have long been a focus of studies of Uranus and Neptune \cp{Hubbard1991,Fortney2011a}.

One possible solution to the composition problem is to obtain multi-wavelength transmission spectra, as has been done for GJ1214b \cp{Bean2011, Desert2011, Croll2011}. Since hydrogen-rich atmospheres have much larger scale heights at a given temperature, near infrared water and methane absorption features will be much more prominent for planets with hydrogen/helium envelopes \cp{Kempton2009,Kempton2010}. Unfortunately, these observations are extremely time intensive and even then the possible presence of clouds can make their interpretation difficult.  Even worse, nearly all the systems found by \textit{Kepler} are too faint for these observations with current telescopes.

An alternative is to develop models of the formation and evolution of low-mass planets to try and predict what compositions can form and how those compositions change as a planet evolves. In particular, hydrodynamic mass loss from extreme ultra-violet (XUV) heating can remove large amounts of hydrogen/helium from highly irradiated LMLD planets. Models of XUV driven mass loss were first developed to study water loss from early Venus \cp{Hunten1982, Kasting1983}, and hydrogen loss from the early Earth \cp{Sekiya1980, Watson1981}. These kinds of models have since been developed to study mass loss from hot Jupiters \cp[e.g.,][]{Lammer2003, Yelle2004, Murray-Clay2009, Ehrenreich2011, Owen2012}, where there is strong evidence that atmospheric escape is an important physical process \cp{Vidal-Madjar2004, Davis2009, Lecavelier2010, Lecavelier2012}.

In Sections \ref{masslosssec}, \ref{watersec}, and \ref{formationsec} we show that energy-limited hydrodynamic mass loss models, coupled with models of thermal evolution and contraction, can distinguish between water-poor super-Earth and water-rich sub-Neptune scenarios in Kepler-11. Moreover, these models make powerful predictions for the density distribution of the entire population of LMLD transiting planets. In particular, observations show that there is threshold in the bulk density - incident flux distribution above which there are no LMLD planets. In Section \ref{fdsec} we examine this threshold and show how it can by reproduced using our thermal evolution models coupled with standard hydrodynamic mass loss prescriptions. Finally, in Section \ref{constraintsec} we explore how this threshold can be used to obtain important constraints on planets without measured densities:  We constrain the maximum radii of non-transiting radial velocity planets, and the minimum masses of \textit{Kepler} candidates.

\section{Our Model}

\subsection{Planet Structure}\label{intsec}
We have built on previous work in \ct{Fortney2007} and \ct{Nettelmann2011} to develop models of the thermal evolution of LMLD planets.  To simplify what is undoubtedly a complex interior structure for real planets, we construct model planets with well-defined layers.  Low-mass planets are likely to have a significant fraction of their mass in iron and silicate rocks. For simplicity, we assume that these materials are contained in a isothermal rocky core with Earth-like proportions of 2/3 silicate rock and 1/3 iron. For the rock, we use the ANEOS \cp{Thompson1990} olivine equation of state (EOS); while for the iron, we use the SESAME 2140 Fe EOS \cp{Lyon1992}.

On top of this rock/iron core we then attach an interior adiabat. The composition of this adiabat depends on the planet model being considered. For this work, we consider three classes of LMLD planets: rocky super-Earths with H/He envelopes, water-worlds that have pure water envelopes, and sub-Neptunes with a water layer in between the core and the upper H/He layer. For the water-rich sub-Neptune models we assume that this intermediate water-layer has the same mass as the rock/iron core. We choose this value because it is comparable to the water to rock ratio need to fit Kepler-11b as a water-world. This allows us to explore the proposition that all five Kepler-11 planets started out with similar compositions, but that mass loss has subsequently distinguished them. For hydrogen/helium we use the \ct{Saumon1995} EOS. Meanwhile for water we use the ab-initio H2O-REOS EOS developed by \ct{Nettelmann2008} and \ct{French2009}, which was recently confirmed up to 7 Mbar in laboratory experiments \cp{Knudson2012}. 

In the Kepler-11 system, our models predict that water will be in the vapor, molecular fluid, and the ionic fluid phases. The interiors are too hot for high pressure ice phases. Finally, we model the radiative upper atmosphere by assuming that the planet becomes isothermal at pressures where the adiabat is cooler than the planet's equilibrium temperature, assuming 20\% Bond albedo and uniform re-radiation. We then calculate the radius at 10 mbar which we take to be the transiting radius.

We connect the different layers of our models models by requiring that pressure and temperature are continuous across boundaries. We then solve for the interior structure assuming hydrostatic equilibrium. A given model is defined by its mass, composition (i.e., the relative proportions in H/He, water, and the rock/iron core) and the entropy of its interior H/He adiabat. By tracking changes in composition and entropy we can then connect these models in time and study the thermal and structural evolution of a given planet.

\subsection{Thermal Evolution}
\label{tesec}

In order to obtain precise constraints on composition, it is important to fully model how a planet cools and contracts due to thermal evolution. Models that only compute an instantaneous structure \cp{Rogers2010b} by necessity must vary the intrinsic luminosity of the planet over several orders of magnitude, which can introduce large uncertainties in the current composition. Obtaining precise constraints from thermal evolution is essential when considering mass loss, since mass loss histories are highly sensitive to uncertainties in the current composition. Moreover, since mass loss depends strongly on planetary radius (to the third power), the mass loss and thermal histories are inextricably linked.

Modeling this contraction requires a detailed understanding of a planet's energy budget. By tracking the net luminosity of a planet, we know how the specific entropy $S$, (i.e., the entropy per unit mass) of the interior adiabat changes with time. For a given mass and composition, this adiabat then defines the planet's structure and so we can track the planet's total radius as the model cools and contracts with time. Equation (\ref{thermaleq}) shows the energy budget for our models and how this relates to the change in entropy $dS/dt$.

\begin{equation}\label{thermaleq}
\int_{M_{\mathrm{core}}}^{M_{\mathrm{p}}} dm \frac{T dS}{dt} = - L_{\mathrm{int}} + L_{\mathrm{radio}} - c_{\mathrm{v}} M_{\mathrm{core}}  \frac{dT_{\mathrm{core}}}{dt}
\end{equation}

The left hand side shows the rate of change of the thermal energy of the interior adiabat. Positive terms on the right hand side represent energy sources that heat and inflate a planet, while negative terms represent energy losses that allow a planet to cool and contract. The term $L_{\mathrm{int}} = L_{\mathrm{eff}}-L_{\mathrm{eq}}$ describes the intrinsic luminosity due to radiation from the planet, where $L_{\mathrm{eq}}$ is the planet's luminosity due only to absorbed stellar radiation.

The $L_{\mathrm{radio}}$ term describes heating due to radioactive decay. The important isotopes are $^{235}$U, $^{40}$K, $^{238}$U, and $^{232}$Th. These have half lives of 0.704, 1.27, 4.47, and 14.1 Gyr, respectively. We assume meteoritic abundances given by \ct{Anders1989}. We do not consider the early decay of $^{26}Al$, since we only consider models that are at least 10 Myr after planet formation. The $L_{\mathrm{radio}}$ term has only a minor effect on our models since it is typically an order of magnitude smaller than the other terms in equation (\ref{thermaleq}).

Lastly, there is the $dT_{\mathrm{core}}/dt$ term, which represents the delay in cooling due to the thermal inertia of the rocky core. As the interior adiabat cools, the core isotherm must also cool, as $T_{\mathrm{core}}$ equals the temperature at the bottom of the adiabat. When the core makes up a large fraction of the planet's mass, this can significantly slow down the planet's rate of contraction. We assume a core heat capacity of $c_v = 0.5-1.0$ $\mathrm{J \, K^{-1} \, g^{-1}}$ \cp{Alfe2002, Guillot1995, Valencia2010} as in \ct{Nettelmann2011}. This range covers values appropriate for both the cores of the Earth and Jupiter. For our three layer sub-Neptune models we still us the mass in rock and iron for $M_{\mathrm{core}}$, since the water layer is generally too hot for ice phases and so it is assumed to be fully convective.

For a given interior structure, we determine the intrinsic flux from the interior, at given $S$ of the adiabat, via interpolation in a grid of model atmospheres.  The values of $T_{\mathrm{int}}$ (a parametrization of the interior flux), $T_{\mathrm{eq}}$, and $T_{\mathrm{eff}}$ are tabulated on a grid of surface gravity, interior specific entropy, and incident flux for 50$\times$ solar metallicity H/He atmospheres (similar to Neptune). This corresponds to a metal mass fraction of $Z \approx 0.35$ and a mean molecular weight of $\mu \approx 3.5$ $\mathrm{g \, mol^{-1} }$. The grid is the same as that described in \ct{Nettelmann2011} for LMLD planet GJ 1214b, where a more detailed description can be found.  Here we do expand on that grid to now include a range of incident fluxes, as was done for giant planets in \ct{Fortney2007}.

In choosing the initial entropy for our evolution model, we assume a ``hot start'' for model; i.e., we start the models out with a large initial entropy. When then allow the models to cool and contract until either 10 Myr or 100 Myr which is when we begin the coupled thermal and mass loss evolution. This is a common but important assumption. However, in general our thermal evolution models are insensitive to the initial entropy choice by $\sim100$ Myr as in \ct{Marley2007}. As a result, we present results at both 10 and 100 Myr. Moreover, to gain confidence in our 10 Myr models we examined the effect of starting those models with a lower initial entropy. Specifically, we ran models in which we started the 10 Myr with the entropies found at 100 Myr. This allowed us to separate the effect of the stellar XUV evolution, from any ``hot start'' vs. ``cold start'' uncertainties. Future progress in modeling the formation of water-rich sub-Neptune planets \cp[e.g.,][]{Rogers2011} may allow for an assessment of the most realistic initial specific entropies.

\subsection{XUV-Driven Mass Loss}
\label{masslosssec}

Close-in planets like those in Kepler-11 are highly irradiated by extreme ultraviolet (EUV) and x-ray photons. These photons photoionize atomic hydrogen high in a planet's atmosphere, which in turn produces significant heating \cp{Hunten1982}. If this heating is large enough, it can generate a hydrodynamic wind that is capable of removing significant mass, potentially including heavier elements as well \cp{Kasting1983}. We couple this XUV driven mass loss to our thermal evolution models following the approach of \ct{Jackson2010} and \ct{Valencia2010}, which explored possible mass loss histories for CoRoT-7b \cp{Leger2009,Queloz2009}. Similar approaches have also been used to study the coupled evolution of hot Jupiters \cp[e.g.,][]{Baraffe2004, Baraffe2005, Hubbard2007a, Hubbard2007b} and hot Neptunes \cp{Baraffe2006}.

A common approach to estimate the mass loss rate is to assume that some fixed fraction of the XUV energy incident on a planet is converted into heat that does work on the atmosphere to remove mass. This is known as the energy-limited approximation \cp{Watson1981} and allows a relatively simple analytic description of mass loss rates.

\begin{equation}\label{masslosseq}
\dot{M}_{\mathrm{e-lim}} \approx \frac{\epsilon \pi F_{\mathrm{XUV}} R_{\mathrm{XUV}}^3}{G M_{\mathrm{p}} K_{\mathrm{tide}}}
\end{equation}
\begin{equation}
K_{\mathrm{tide}} = (1 - \frac{3}{2 \xi} + \frac{1}{2 \xi^3})
\end{equation}
\begin{equation}
\xi = \frac{R_{\mathrm{Hill}}}{R_{\mathrm{XUV}}}
\end{equation}

Equation (\ref{masslosseq}) describes our estimate of the mass loss rate based on the formulation from \cp{Erkaev2007}. $F_{\mathrm{XUV}}$ is the total flux between $1-1200$ \AA, which is given by Ribas(2005) for Sun-like stars. For stars older than 100 Myr, Ribas found that at 1 AU $F_{\mathrm{XUV}} = 29.7 \tau^{-1.23} \: \mathrm{erg \, s^{-1} \, cm^{-2}}$, where $\tau$ is the age of the star in Gyr. Using this power law, we scale the XUV flux to the appropriate age and semi-major axis for each planet in our models. Although Ribas only targeted Sun-like stars, \ct{Sanz-Forcada2010} found similar results for a wide range of stellar types from M3 to F7. Hereafter, we will simply refer to the entire $1-1200$ \AA $;$ spectrum as XUV.

$R_{\mathrm{XUV}}$ is the planetary radius at which the atmosphere becomes optically thick to XUV photons, which \ct{Murray-Clay2009} find occurs at pressures around a nanobar, in the hot Jupiter context. For our work, we assume that the atmosphere is isothermal between the optical and XUV photospheres. This neglects heating from photo-disassociation, which should occur around a $\mathrm{\mu bar}$ \cp{Kempton2012a}. However, this effect should be relatively small and if anything will lead to slight underestimate of the mass loss rate. We vary pressure of the XUV photosphere from 0.1 nbar to 10 nbar to include the uncertainty in the structure of the XUV photosphere. For H/He atmospheres on LMLD planets, the nbar radius is typically 10-20\% larger than the optical photosphere. $K_{\mathrm{tide}}$ is a correction factor that accounts for the fact that mass only needs to reach the Hill radius to escape \cp{Erkaev2007}. For planets like Kepler-11b today this correction factor increases the mass loss rate by $\sim$ 10\%, however at early times it can increase the rate by as much as a factor of 2. 

Finally, $\epsilon$ is an efficiency factor that parametrizes the fraction of the incident XUV flux that is converted into usable work. This efficiency is set by radiative cooling, especially via Lyman $\alpha$, and can depend on the level of incident flux \cp{Murray-Clay2009}. Kepler-11 is a $8\pm2$ Gyr old Sun-like star. Using the power law from Ribas et al., this implies that current XUV flux at Kepler-11f is $\approx 37 \: \mathrm{erg \, s^{-1} \, cm^{-2}}$. Similarly, when Kepler-11 was 100 Myr old, the flux at Kepler-11b was $\approx 6\times10^4 \: \mathrm{erg \, s^{-1} \, cm^{-2}}$. \ct{Murray-Clay2009} found that at XUV fluxes over $10^5 \: \mathrm{erg \, s^{-1} \, cm^{-2}}$, relevant for many hot Jupiters, mass loss becomes radiation/recombination-limited and highly inefficient. However, at the lower XUV fluxes relevant for the Kepler-11 system mass loss is roughly linear with $F_{\mathrm{XUV}}$ and has efficiencies $\sim 0.1-0.3$. For this work, we assume a default efficiency of $\epsilon=0.1\pm^{0.1}_{0.05}$, although we do examine the effects of lower efficiencies.  While we predominantly investigate the loss of H/He envelopes, in some limited cases for Kepler-11b, we also assume this holds for steam envelope loss.

\begin{deluxetable*}{ccccccccc}[h!]
  \footnotesize
  \tablecaption{Current Mass and Composition}
  \tablewidth{0pt}
  \tablehead{
   \colhead{Planet} & \colhead{Current Mass ($M_{\mathrm{\oplus}}$)} & \colhead{\% H/He} & \colhead{\% Water} & \colhead{\% H/He 3-layer}
}
  \startdata
   	Kepler-11b &  $4.3\pm^{2.2}_{2.0}$   &  $0.3\pm^{1.1}_{0.25}\%$  &  $40\pm^{41}_{29}\%$  &  n/a  
  \\\\
    	Kepler-11c &  $13.5\pm^{4.8}_{6.1}$   & $4.6\pm^{2.7}_{2.3}\%$  &  n/a  &  $0.3\pm^{1.0}_{0.1}\%$ 
  \\\\
	Kepler-11d &  $6.1\pm^{3.1}_{1.7}$   &  $8.2\pm^{2.7}_{2.4}\%$  &  n/a  &  $1.3\pm^{0.9}_{0.8}\%$ 
  \\\\
 	Kepler-11e &  $8.4\pm^{2.5}_{1.9}$   &  $17.2\pm^{4.1}_{4.2}\%$  &  n/a  &  $5.5\pm^{2.3}_{3.0}\%$   
  \\\\
 	Kepler-11f &  $2.3\pm^{2.2}_{1.2}$   &  $4.1\pm^{1.8}_{1.5}\%$  &  n/a &  $0.4\pm^{0.6}_{0.2}\%$    
  \\\\
  \enddata
  \label{currenttab}
  \tablecomments{Present day masses and compositions for Kepler-11 for three classes of models. The first two columns list each planet along with the observed mass derived from TTV in \ct{Lissauer2011a}. The third column lists the current H/He fractions predicted by our thermal evolution models assuming a two layer H/He on rock/iron or water-poor ``super-Earth'' model. The fourth column shows the predicted water fraction for a two layer steam on rock/iron model or "water-world." This is only applicable to 11b as the other are all less dense than pure water. The final column lists the predicted fraction of H/He for three layer "sub-Neptune'' models with equal mass in the rock and water layer. This scenario is not applicable to 11b since it only needs 40\% water to match its radius.}
\end{deluxetable*}

One important implication from equation (\ref{masslosseq}) is that mass loss rates are much higher when planets are young. This is due to two reasons.  Planetary radii are considerably larger due to residual heat from formation. Moreover, at 100 Myr $F_{\mathrm{XUV}}$ was $\approx 500$ times higher than it is currently \cp{Ribas2005}. As a result, most of the mass loss happens in a planet's first Gyr. Thus although a planet's envelope may be stable today, its composition may have changed significantly since formation. Likewise, a considerable amount of mass will be lost between the end of planet formation at $\sim10$ Myr \cp{Calvet2002} and 100 Myr. Following the x-ray observations of \ct{Jackson2012}, we assume that at ages younger than 100 Myr the stellar XUV flux saturates and is constant at the 100 Myr value. Unfortunately, the observations for 10-100 Myr do not cover the EUV (100-1200 \AA) part of the spectrum, so there is some uncertainty as to whether this saturation age is uniform across the entire XUV spectrum. Nonetheless, our constraints on the formation of Kepler-11 in sections \ref{sesec} and \ref{formationsec} come from the lower limits we are able to place on the initial compositions. Assuming that the EUV saturates along with the x-rays is conservative assumption in terms of the amount of mass that is lost.

In general, models of LMLD planets that assume H/He envelopes today will predict much larger mass loss histories then models that assume steam envelopes. Partly, this is because the lower mean molecular weight of hydrogen. Mostly, however, it is because when we integrate the compositions back in time from the present, the addition of a small amount of H/He has much larger impact on a planet's radius than a small amount of water. A larger radius in the past in turn means a higher mass loss rate; and so the integrated mass loss history becomes much more substantial for H/He envelopes.

\section{Application to Kepler-11}

\subsection{Current Compositions from Thermal Evolution}

The first step in trying to understand the formation and history of a planetary system is to identify the possible current compositions for each of the planets in the absence of any mass loss. This then gives us estimates for the current masses of each planet's core, which we then use as the starting point for all of our calculations with mass loss.

\begin{figure*}[h!] 
  \begin{center}
    \includegraphics[width=6.0in,height=4.3in]{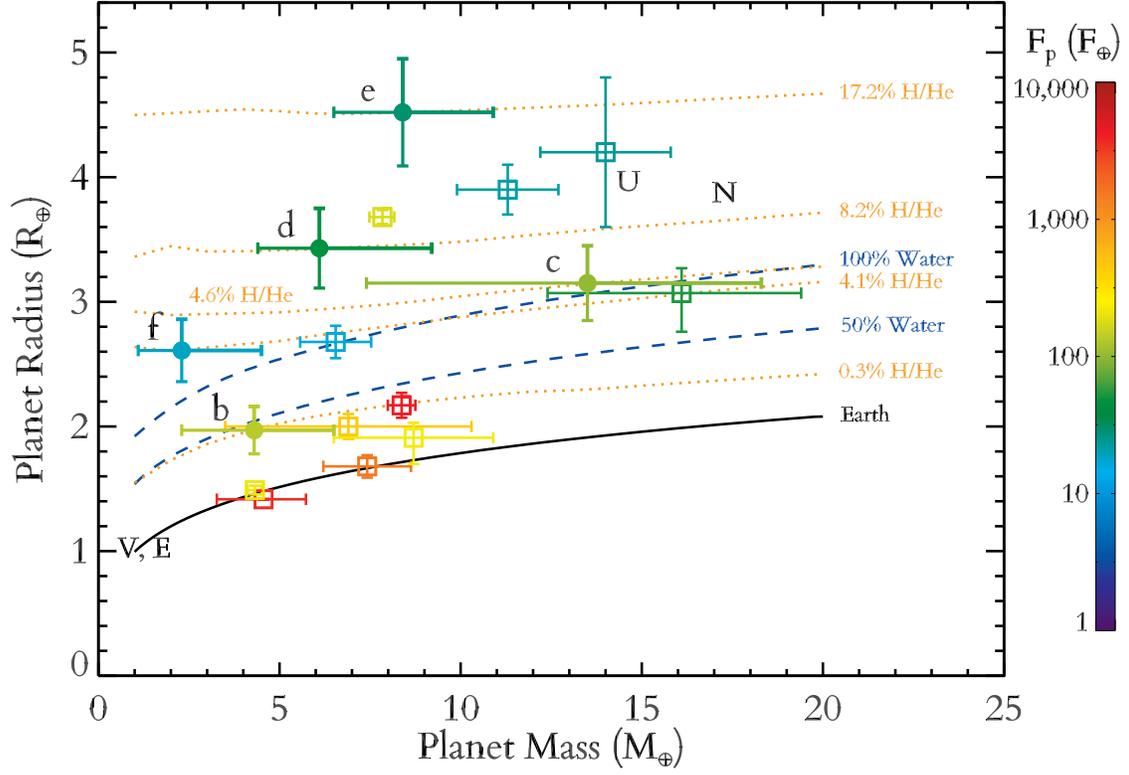}
  \end{center}
  \caption{Radius vs. mass for transiting exoplanets with measured masses, along with curves for different compositions. Planets are color-coded by the incident bolometric flux they receive. Kepler-11 planets are shown by filled circles with letters indicating each planet. Other known exoplanets in this mass and radius range are shown by open squares. Solar system planets Earth, Venus, Uranus, and Neptune are shown by black letters. The solid black curve is for a Earth-like composition with 2/3 rock and 1/3 iron. All other curves use full thermal evolution calculations, assuming a volatile envelope atop a earth like core. The dashed blue curves are for 50\% and 100\% water by mass. The dotted orange curves are for H/He envelopes at 8 Gyr; each one is tailored to match a Kepler-11 planet and is computed at the appropriate flux and for that planet.\label{mrfig}}
\end{figure*}

Figure \ref{mrfig} shows the Kepler-11 planets in a mass-radius diagram along with curves for different possible compositions. For all planets, we color-code by the incident bolometric flux they receive. The Kepler-11 planets are shown by filled circles with identifying letters next to each one. The other known transiting exoplanets in this mass and radius range are shown by the open squares. In order of increasing radius, these are Kepler-10b \cp{Batalha2011}, Kepler-36b \cp{Carter2012}, CoRoT-7b \cp{Leger2009,Queloz2009,Hatzes2011}, Kepler-20b \cp{Fressin2011,Gautier2011}, Kepler-18b \cp{Cochran2011}, 55 Cancri e \cp{Winn2011,Demory2011}, GJ 1214b \cp{Charbonneau2009}, Kepler-36c \cp{Carter2012}, Kepler-30b \cp{Fabrycky2012, Sanchis-Ojeda2012}, and GJ 3470b \cp{Bonfils2012}. Lastly, the open triangles show the four planets in our own solar system that fall in this range: Venus, Earth, Uranus, and Neptune.

The curves show various possible compositions. The solid black curve shows a standard Earth-like composition with 2/3 rock and 1/3 iron as described in Section \ref{intsec}. The other curves show compositions with thick water or H/He envelopes atop an Earth-like core. These curves include thermal evolution without mass loss to 8 Gyr, the age of Kepler-11. The blue dashed curves show the results for 50\% and 100\% water-worlds computed at $T_{\mathrm{eq}}=700$ K, approximately the average temperature of the five inaner planets. Likewise, the dotted orange curves show the results for H/He envelopes; however, here each curve is tailored to match a specific Kepler-11 planet and is computed at the flux of that planet. These fits are listed in greater detail in table \ref{currenttab}. Here we list the mass of each planet taken from \ct{Lissauer2011a}; the H/He fractions needed to match each planet's current radius for a water-poor super-Earth model; the water fraction needed to match Kepler-11b as a water-world; and the H/He fractions needed to fit Kepler-11c, d, e, and f as sub-Neptunes with an intermediate water layer, as described in Section \ref{intsec}. As described in section \ref{sesec}, we varied the planetary albedo, the heat capacity of the rocky core, and the observed mass, radius, current age, and incident flux.

Figure \ref{mrfig} and table \ref{currenttab} clearly show the degeneracy between various compositions that we are attempting to untangle. There are now four planets including Kepler-11b that can easily be fit either as water-worlds or as water-rich sub-Neptunes with $<2\%$ of their mass in H/He.  However, it is worth looking closer at Kepler-11b in particular. It is this the only planet in the system which does not require any hydrogen or helium to match its current radius, although it must have some sort of volatile envelope. Moreover, it is also the most irradiated and it is fairly low gravity. As a result, adding a small amount of hydrogen to its current composition has a large impact on the bulk density, which in turn makes the planet more vulnerable to mass loss, as seen in Eq. (\ref{masslosseq}).  A clearer picture for this planet emerges when including XUV driven mass-loss and relatively strong constraints from formation models discussed in Section \ref{formationsec}. Thus, if there is hope of using mass loss to constrain the composition and formation of the system, it likely lies with Kepler-11b.

\subsection{Mass Loss for a Super-Earth Scenario}
\label{sesec}

\begin{deluxetable*}{cccccccccc}[h!]
  \footnotesize
  \tablecaption{Results from Mass Loss: Mass and Composition for Super-Earths}
  \tablewidth{0pt}
  \tablehead{
   \colhead{Planet} & \colhead{Mass 100 Myr ($M_{\mathrm{\oplus}}$)} & \colhead{\% H/He 100 Myr} & \colhead{Mass 10 Myr ($M_{\mathrm{\oplus}}$)} & \colhead{\% H/He 10 Myr} 
}
  \startdata
   	Kepler-11b &    $34.6\pm^{6.5}_{28.2}$  &  $87.6\pm^{6.6}_{85.4}\%$  &  $44.8\pm^{9.7}_{10.1}$  &  $90.4\pm^{5.1}_{8.2}\%$  
  \\\\
    	Kepler-11c &    $13.7\pm^{4.7}_{5.8}$   &  $6.0\pm^{5.0}_{3.2}\%$  &  $14.2\pm^{4.3}_{3.1}$  &  $9.1\pm^{28}_{7.3}\%$ 
  \\\\
	Kepler-11d &    $6.7\pm^{2.8}_{0.6}$    &  $16.5\pm^{22}_{8.5}\%$  &  $7.8\pm^{12.8}_{0.8}$  &  $28\pm^{56}_{17}\%$ 
  \\\\
 	Kepler-11e &    $8.8\pm^{2.3}_{1.6}$    &  $21.2\pm^{6.0}_{3.2}\%$  &  $9.7\pm^{2.5}_{1.9}$  &  $28.1\pm^{10.5}_{7.7}\%$   
  \\\\
 	Kepler-11f &    $3.1\pm^{5.2}_{0.2}$    &  $29\pm^{58}_{24}\%$  &  $3.4\pm^{6.6}_{0.4}$  &  $35\pm^{57}_{25}\%$    
  \\\\
  \enddata
  \label{masslosstab}
  \tablecomments{Masses and H/He fractions predicted by coupled mass loss and thermal evolution models at 100 and 10 Myr, assuming all five planets are water-poor super-Earths. The large error bars on some compositions are due mostly to uncertainties in the current masses from TTV. The 10 Myr values are subject to some model uncertainties as discussed in sections \ref{tesec} and \ref{masslosssec}. Kepler-11b is extremely vulnerable to H/He mass loss and would have to start off implausibly massive to retain a small H/He envelope today. For the best-fit masses, Kepler-11c is less vulnerable to mass loss due to its massive core. However, Kepler-11c, d, e, and f are all consistent with have formed with $\sim$30\% H/He.}
\end{deluxetable*}

Now that we have estimates for the present day compositions, we will begin considering the effects of mass loss. We will compute mass loss histories that when evolved to the present day, match the current mass and composition. This then tells us what the mass would have to be in the past to result in the current mass and composition.  As discussed in Section \ref{masslosssec}, there is uncertainty in stellar XUV fluxes ages younger than 100 Myr; as a result, we will present results both at 10 Myr and 100 Myr after planet formation.

\begin{figure*}[h!] 
  \begin{center}
    \includegraphics[width=5.in,height=3.57in]{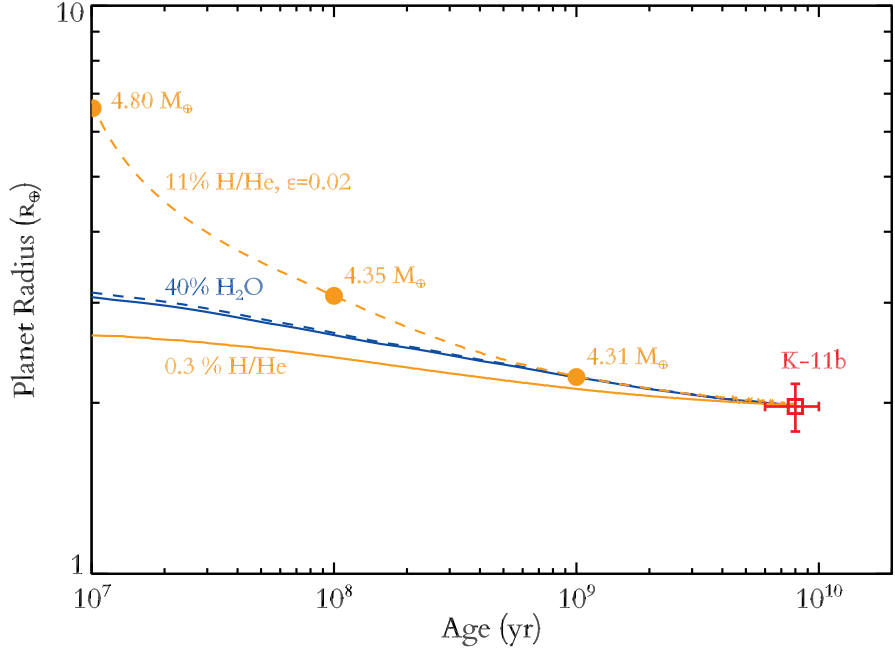}
  \end{center}
  \caption{Radius vs. time for four example model runs that match the present day mass and radius of Kepler-11b. The blue curves show water-world models, while the orange curves show water-poor super-Earth models. Dashed lines are with mass loss, while solid are without. Both water-world models and the water-poor super-Earth model without mass loss show very similar cooling curves. Even with our standard efficiency of 10\% the water models undergo only minor mass loss. Meanwhile, even with an efficiency 4$\times$ smaller the H/He model undergoes substantial mass loss. This model is initially 14\% H/He and 5.0 $M_{\mathrm{\oplus}}$. We have marked the masses for the H/He with mass loss model at 10 Myr, 100 Myr, and 1 Gyr. This also shows the large impact on radius that even a modest (compared to Figure \ref{runawayfig}) H/He envelope can have. \label{k11bfig}}
\end{figure*}

First we will consider water-poor super-Earth models for each planet, which have H/He envelopes atop Earth-like rocky cores. As discussed in Section \ref{masslosssec}, H/He envelopes are particularly susceptible to mass loss. As an example, Figure \ref{k11bfig} shows four possible cooling histories for Kepler-11b. The solid lines show thermal evolution without any mass loss while the dashed lines include mass loss. The orange curves are for water-poor super-Earth models, while the blue curves show water-world models. The red cross shows the current radius and age of Kepler-11b. These curves illustrate the impacts of both thermal evolution and mass loss on the radius of a low-mass planet. The water-world models require that 40\% of the current mass must be in water to match the current radius. Assuming our standard efficiency $\epsilon=0.1$, implies an initial composition of 43\% water at 10 Myr. This illustrates the relative stability of water envelopes. On the other hand, the dashed orange curve shows the vulnerability of H/He layers. Here we have assumed a efficiency 5$\times$ lower $\epsilon=0.02$ and yet more mass is lost than in the water-world scenario. Even at this low efficiency, Kepler-11b would have to initially be 11\% H/He and 4.8 $M_{\mathrm{\oplus}}$ to retain the 0.3\% needed to match the current radius. This also shows the large increase in radius that can result from even a relatively modest increase in the H/He mass.

\begin{figure*}[h!] 
  \begin{center}
    \includegraphics[width=5.in,height=6.43in]{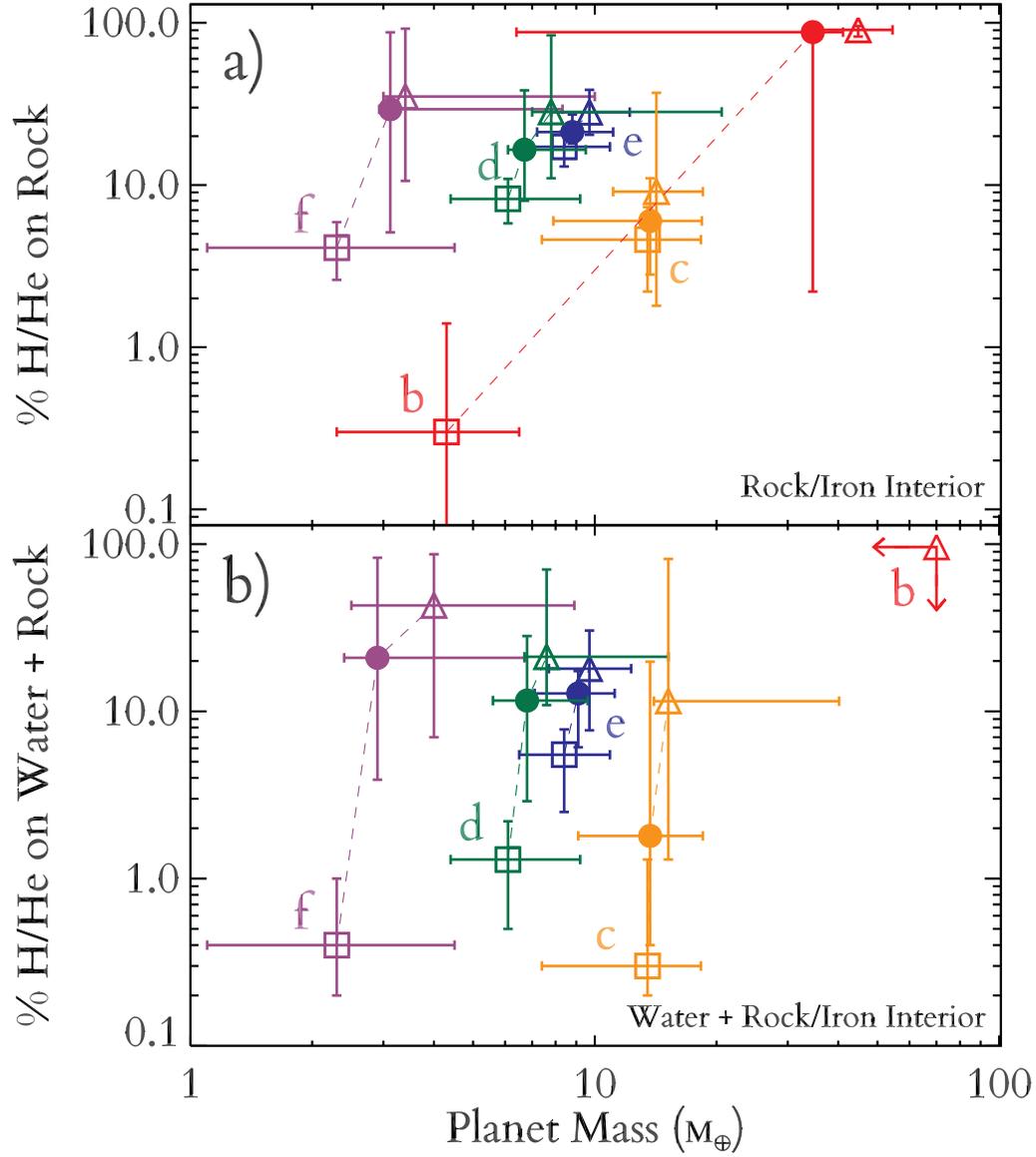}
  \end{center}
  \caption{Composition vs. mass for models of Kepler-11 with mass loss. Panel a) shows the results for water-poor super-Earth models with a H/He envelope atop a rock/iron core. Panel b) shows the results for water-rich sub-Neptune models that also have a thick water layer in between. Each point shows the \% H/He and mass predicted by our thermal evolution and mass loss models at a given time. Each color indicates a particular planet as identified by the letters and connected by dashed lines. The open squares show the present day mass and composition as listed in table \ref{currenttab}. The filled circles show the results at 100 Myr and the open triangles show the results at 10 Myr as listed in tables \ref{masslosstab} and \ref{watertab}. Kepler-11b is extremely vulnerable to H/He mass loss, and would have been up to $\sim90\%$ H/He if it formed as a water-poor super-Earth. Kepler-11c is the least vulnerable due to its larger core mass, and d, e, and f are intermediate. All five planets are consistent with having initially been water-rich sub-Neptunes with comparable amounts of rock and water and $\sim$20\% of their mass in H/He. \label{lossfig}}
\end{figure*}

Table \ref{masslosstab} summarizes the results for Kepler-11 b-f for the water-poor super-Earth scenario. We list the masses predicted by our models when the planets were 10 and 100 Myr old. In addition, we list the fraction of the planets' masses in the H/He envelope at each age. These results are further illustrated in Figure \ref{lossfig}a. Here we have plotted the mass and H/He fraction for each planet at 10 Myr, 100 Myr, and today. Each color corresponds to a particular planet with the squares indicating the current masses and compositions, the circles the results at 100 Myr, and the triangles the results at 10 Myr. In order to calculate the uncertainty on these results, we varied the mass loss efficiency $\epsilon$ from 5-20\% and varied the XUV photosphere from 0.1-10 nbar. Likewise, we varied the planetary Bond albedo from 0-0.80 and varied the heat capacity of the rocky core from 0.5-1.0 $\mathrm{J \, g^{-1} \, K^{-1}}$. Also, as discussed in section \ref{tesec}, we varied the initial entropy for the 10 Myr models, to account for undertainties in ``hot-start'' vs. ``cold-start''. Finally, we factored in the observed uncertainties in mass, radius, and incident flux.

\begin{figure*}[h!] 
  \begin{center}
    \includegraphics[width=5.35in,height=3.57in]{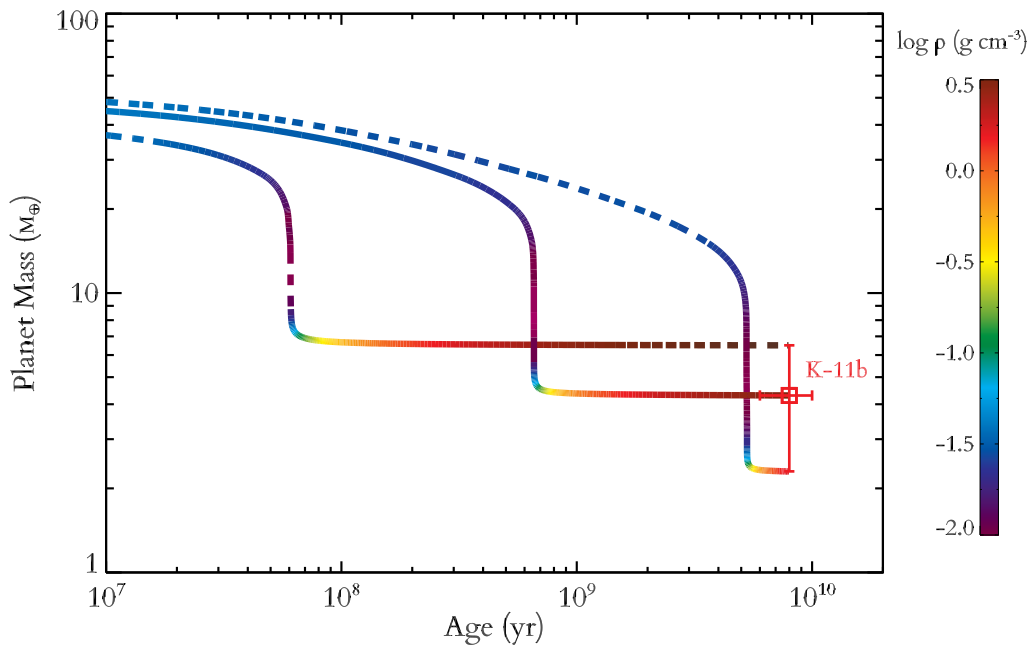}
  \end{center}
  \caption{Mass vs. time with mass loss for three model runs that match the present day mass along with its 1$\sigma$ range for Kepler-11b. All three models assume a water-poor super-Earth composition that is 0.3\% H/He today. The curves are color-coded by log density. The solid line corresponds to the best fit current mass from TTV; the dashed lines correspond to the 1$\sigma$ bounds. This demonstrates several features described in the text. The initial mass is actually lower if Kepler-11b is more massive today due to a correspondingly more massive core. There is a period of runaway mass loss during which the density actually declines slightly, and the timing of this period depends strongly on the mass of the rocky core. \label{runawayfig}}
\end{figure*}

Clearly, Kepler-11b is vulnerable to extreme mass loss if it has a H/He envelope atop a rock/iron core. Although less than 1\% H/He today, if it is a water-poor super-Earth it could have been have over $\sim90\%$ H/He in the past. At 10 Myr, its mass would have been $45\pm10$ $M_{\mathrm{\oplus}}$, an order of magnitude higher than the current value. Kepler-11b is able to undergo such extreme mass loss because its high XUV flux and the low mass of its rocky core put it in a regime where it is possible to enter a type of runaway mass loss. This happens when the mass loss timescale is significantly shorter then the cooling timescale. After the planet initially loses mass it has an interior adiabat and rocky core that are significantly hotter than would otherwise be expected for a planet of its mass and age. This is because the interior still remembers when the planet was more massive and has not had sufficient time to cool. As a result, the planet will stay inflated for some time and the density stays roughly constant and can actually decrease. A similar effect was seen by \ct{Baraffe2004} when they studied coupled thermal evolution and mass loss models for core-less hot Jupiters. We find that this process generally shuts off once the composition drops below $\sim$20\% H/He. At that point the presence of the core forces the total radius to shrink even if the planet is unable to cool efficiently. Figure \ref{runawayfig} shows this process as Kepler-11b loses mass for three different values of its current mass and therefore its core mass. The curves correspond to the best fit mass from transit-timing as well as the 1$\sigma$ error bars. This shows that the timing of this runaway loss event depends strongly on the mass of the rock/iron core. 

Super-Earth models of Kepler-11b are unusual in that they are subject to tremendous mass loss and yet they retain a small amount of H/He today. Typically models that start out $\sim90\%$ H/He either experience runaway mass loss and lose their H/He envelopes completely, or they never enter the runaway regime and remain over 50\% H/He. The uncertainty in the initial composition of Kepler-11b is due to uncertainty in its TTV mass. At a given current mass, the range of Kepler-11b models that will retain an envelope that is $<1\%$ H/He is extremely narrow. In this sense, the current composition of Kepler-11b requires a rare set of initial conditions if it is a water-poor super-Earth.

\begin{deluxetable*}{cccccccccc}[h!]
  \footnotesize
  \tablecaption{Results from Mass Loss: Water-Worlds and Sub-Neptunes}
  \tablewidth{0pt}
  \tablehead{
  \colhead{Planet} & \colhead{Mass 100 Myr ($M_{\mathrm{\oplus}}$)} & \colhead{Composition 100 Myr} & \colhead{Mass 10 Myr ($M_{\mathrm{\oplus}}$)} & \colhead{Composition 10 Myr} 
}
  \startdata
  Kepler-11b &    $4.4\pm^{2.2}_{2.0}$   &  $41\pm^{39}_{28}\%$  &  $4.5\pm^{2.1}_{1.8}$  &  $43\pm^{38}_{29}\%$ 
  \\\\
    	Kepler-11c &    $13.7\pm^{4.8}_{4.6}$   &  $1.8\pm^{18}_{1.4}\%$  &  $15.2\pm^{25}_{1.2}$  &  $12\pm^{70}_{10}\%$ 
  \\\\
	Kepler-11d &    $6.8\pm^{2.8}_{1.2}$   &  $11.6\pm^{17}_{8.7}\%$  &  $7.6\pm^{7.6}_{0.9}$  &  $21\pm^{49}_{10}\%$ 
  \\\\
 	Kepler-11e &    $9.1\pm^{2.1}_{2.0}$   &  $12.8\pm^{4.6}_{6.7}\%$  &  $9.7\pm^{2.6}_{2.0}$  &  $18\pm^{12}_{10}\%$   
  \\\\
 	Kepler-11f &    $2.9\pm^{3.8}_{0.5}$   &  $21\pm^{62}_{17}\%$  &  $4.0\pm^{4.9}_{1.5}$  &  $43\pm^{44}_{36}\%$    
  \\\\
  \enddata 
  \label{watertab}
  \tablecomments{Masses and volatile fractions predicted by coupled mass loss and thermal evolution models at 100 and 10 Myr, assuming Kepler-11c, d, e, and f are water-rich sub-Neptunes and Kepler-11b is a water-world. Thus the compositions listed for Kepler-11b are water fractions, while those for Kepler-11c-f are H/He fractions. Kepler-11c-f are all consistent with having formed as water-rich sub-Neptunes with 20-30\% H/He. Kepler-11b is not vulnerable to mass loss if it has a water envelope; however, it could have easily also formed a water-rich sub-Neptune and lost its H/He envelope.}
\end{deluxetable*}

Counterintuitively, if Kepler-11b is more massive today then its implied mass in the past is actually lower. This is because a higher mass today would imply a more massive core, which would increase the planet's density and decrease its mass loss rate. As a result, a more massive model for Kepler-11b today is less vulnerable to mass loss and so less H/He is needed in the past in order to retain 0.3\% today. At 100 Myr, there is a very large uncertainty in the composition due to the uncertainty in the core mass. However, even if we assume the 1$\sigma$ error bar 6.5 $M_{\mathrm{\oplus}}$, Kepler-11b would still be at least 37 $M_{\mathrm{\oplus}}$ and at least 83\% H/He at 10 Myr. In section \ref{formationsec}, we will compare this to models of in situ formation and show that such a scenario is unlikely.

On the other hand, Kepler-11c is not particularly vulnerable to mass loss, at least using the best fit mass from transit timing, despite having the second highest flux in system. This is because of the relatively large mass of its rocky core; the high gravity means additional H/He has a more modest effect of the planet's radius and therefore on the mass loss rate. In fact, along with the incident XUV flux the mass of the rocky core is the single largest factor that determines whether a given planet will be vulnerable to mass loss. As a result, the dominant sources of uncertainty in our mass loss models are the uncertainties in the masses from TTV. These dominate over all the theoretical uncertainties in the thermal evolution and mass loss models. The uncertainty in planet mass from transit timing is particular large for Kepler-11c. If its mass is close to the 1$\sigma$ low value, then it is possible Kepler-11c has undergone more substantial mass loss similar to Kepler-11d-f. Fortunately, as more quarters of data are processed the mass estimates from TTV will become more precise \cp{Agol2005,Holman2005}. Finally, Kepler-11d, e, and f are modestly vulnerable to mass loss and are consistent with having originated with $\sim 20\%$ H/He at 100 Myr and $\sim 30\%$ H/He at 10 Myr. In Section \ref{hillsec} we will discuss these results in terms of orbital stability.

\subsection{The Water-Rich Scenario}
\label{watersec}

Next we consider a water-rich scenario where the entire system formed beyond the snow line. We assume that Kepler-11c-f are water-rich sub-Neptunes as described in Section \ref{intsec}, while Kepler-11b is currently a water-world. Otherwise the thermal mass loss histories are calculated in the same manner as the water-poor super-Earth scenario. For Kepler-11c-f we calculate the planet mass H/He fraction at 10 and 100 Myr, assuming that only H/He is lost. For Kepler-11b, we examine the vulnerability of both H/He and steam envelopes atop water-rich interiors. The results are summarized in Table \ref{watertab} which list the water fraction for a water-world model of Kepler-11b and the H/He fraction for water-rich sub-Neptune models of Kepler-11c-f. Likewise, the results for c-f are shown in Figure \ref{lossfig}b.

In general, these three layer models are slightly more vulnerable to mass loss than the water-poor super-Earth models presented in section \ref{sesec}. Mostly this is because models with a water layer have hotter interiors that cool more slowly. Since models For example, for Kepler-11c without mass loss the models presented in Table \ref{currenttab}, at 8 Gyr the final entropy in the H/He layer is 6.6 $\mathrm{k_b}$/baryon for the water-rich sub-Neptune model versus 5.8 for the water-poor super-Earth model. The second reason is that counter-intuitively the water-rich sub-Neptune models are slightly more vulnerable to mass loss precisely because they have less of the planet's mass in H/He today. For a planet that has less H/He today, adding a small amount of H/He at the margin has a larger impact on the planet's radius and therefore on the mass loss rate.

For Kepler-11c-f the results are broadly similar to the those for the water-poor super-Earth scenario. Kepler-11c is again the least vulnerable to mass loss; while Kepler-11d is again the most vulnerable of the four planets that we model as water-rich sub-Neptunes. However, all four of these planets are consistent with having been $\sim10-20\%$ H/He at 100 Myr and $\sim20-30\%$ H/He at 10 Myr. 

If Kepler-11b was always a water-world, then mass loss was never important for it. Between 10 Myr and the present, it only drops from 43\% to 40\% water. Moreover, if Kepler-11b was initially a water-rich sub-Neptune similar to the other planets in the system, it could have easily stripped its H/He outer envelope. If we start Kepler-11b at 100 Myr as a water-rich sub-Neptune similar to the other planets with $30\%$ H/He atop 4.3 $M_{\mathrm{\oplus}}$ of rock and water, then assuming $\epsilon=0.1$ the entire H/He envelope will be stripped by 300 Myr. We can set upper limits on the initial mass and H/He fraction of 70 $M_{\mathrm{\oplus}}$ and 94\% if Kepler-11b was originally a water-rich sub-Neptune. These are however strictly upper limits, a H/He layer could have been lost at any time between formation and now. Therefore, all five planets are consistent with a scenario is which they formed as water-rich sub-Neptunes with $\sim10\%$ H/He at 100 Myr and $\sim20\%$ H/He at 10 Myr.

\section{A Mass Loss Threshold for Low-Mass Low-Density Planets}
\label{fdsec}

\begin{figure*}[h!] 
  \begin{center}
    \includegraphics[width=6.in,height=4.3in]{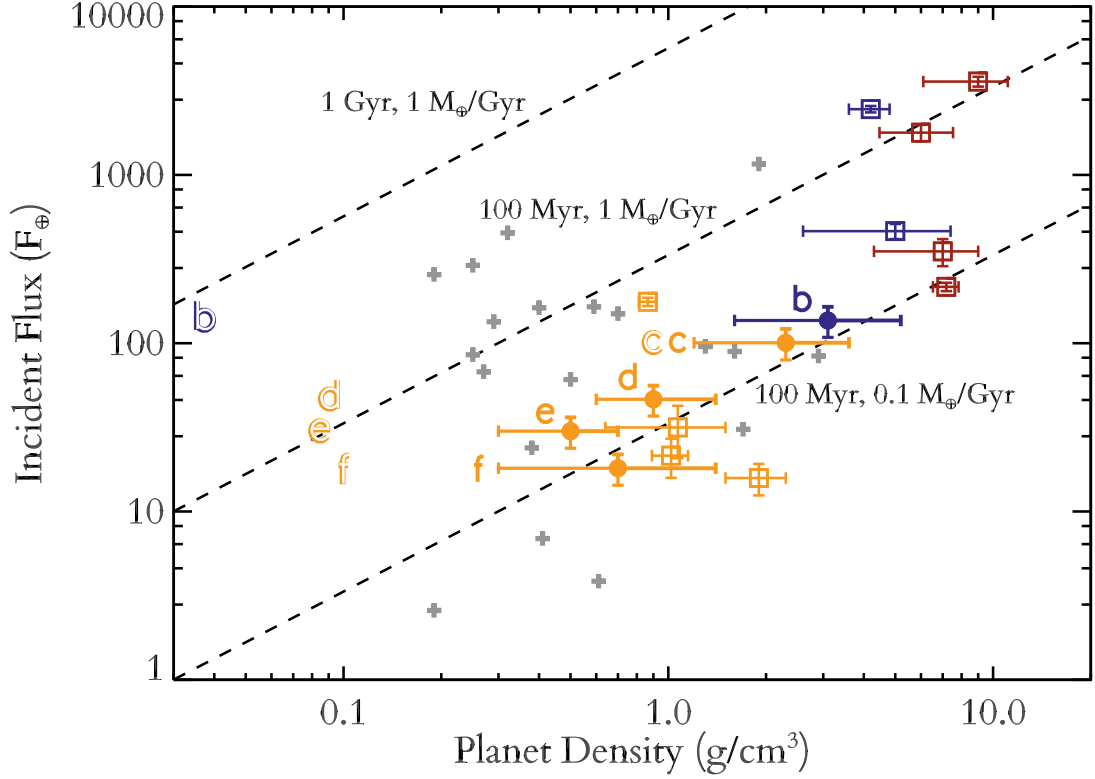}
  \end{center}
  \caption{Bolometric flux at the top of the atmosphere, relative to the flux incident on Earth, vs. planet density. Once again, Kepler-11 planets are shown by filled circles with letters indicating each planet. Open squares show the other extrasolar planets included in Figure \ref{mrfig}. Colors indicate possible compositions. Planets that could be rocky are red, those that could be water-worlds are blue, and those that must have H/He are orange. For comparison, the gray crosses show all other transiting planets with measured masses greater than 15 $M_{\mathrm{\oplus}}$ and less than 100 $M_{\mathrm{\oplus}}$. The dashed black lines show curves of constant mass loss for different mass loss rates and ages, assuming our standard mass loss efficiency of 20\%. Finally, the shaded letters at the left indicate the densities for each Kepler-11 planet at 100 Myr predicted by our mass loss evolution models in Section \ref{sesec}. \label{shorelinefig}}
\end{figure*}

Although Kepler-11 provides a unique case-study, it is essential to explore how mass loss impacts the larger population of LMLD transiting planets. Figure \ref{shorelinefig} shows the bolometric flux these planets receive at the top of their atmospheres vs. their bulk densities. As in Figure \ref{mrfig}, filled circles show the Kepler-11 planets with the letters indicating each planet. Likewise, the open squares show the other transiting exoplanets that are less than 15 Earth masses. For reference, we have also plotted all other transiting planets between 15 and 100 $M_{\mathrm{\oplus}}$ as gray crosses \cp{Wright2011}. The colors indicate possible compositions. All planets with a best-fit mass and radius that lies below a pure rock curve are colored red. These include Kepler-10b, Kepler-36b, CoRoT-7b, and just barely Kepler-20b. Planets that are less dense than pure rock but more dense than pure water, indicating that the could potentially be water-worlds, are colored blue. These include Kepler-11b, Kepler-18b, and 55 Cancri e. Meanwhile those planets that must have a H/He envelope to match their radius are colored orange. These include Kepler-11c, d, e, and f, Kepler-30b, Kepler-36c, GJ 1214b, and GJ 3470b.

The dashed black lines show curves of constant mass loss rate according to equation (\ref{masslosseq}), assuming $\epsilon=0.1$ and $K_{\mathrm{tide}}=1$. These curves are linear in this plot since the instantaneous mass loss rate goes as the flux over the density. Although Figure \ref{shorelinefig} plots the bolometric flux today, we can relate this to an XUV flux at a given time using the \ct{Ribas2005} power law for sun-like stars described in Section \ref{masslosssec}. The curves show the flux today required to lose mass at 1 $M_{\mathrm{\oplus}} \, \mathrm{Gyr^{-1}}$ when the planets were 1 Gyr old and 100 Myr old, along with another curve showing 0.1 $M_{\mathrm{\oplus}} \, \mathrm{Gyr^{-1}}$ at 100 Myr. Since most of the mass loss happens in the first few hundred Myrs, the bottom two curves can roughly be considered as the respective thresholds for mass loss being important and being unimportant for LMLD planets.

One possible explanation of this mass loss threshold is that it caused by XUV driven mass loss from H/He envelopes on low-mass planets. LMLD planets that form above the 100 Myr 1 $M_{\mathrm{\oplus}} \, \mathrm{Gyr^{-1}}$ curve lose mass, increase in density and move to the right until they lie below this threshold. The planets that are left above this line are mostly rocky or at the very least probably do not have H/He envelopes. Planets more massive than $\sim 15$ $M_{\mathrm{\oplus}}$ are not affected since they have a larger reservoir of mass and the loss of a few earth masses of volatiles isn't sufficient to significantly change their bulk density. To illustrate this, we have plotted our predictions for the bulk densities of each of the Kepler-11 planets at 100 Myr, including the effects of both mass loss and thermal evolution. These are indicated by the shadowed letters at the left of Figure \ref{shorelinefig}.


The situation becomes even clearer if we instead we plot flux against mass times density as in Figure \ref{improvedfig}. The timescale for XUV mass loss goes like $\rho M_{\mathrm{p}}/F_{\mathrm{XUV}}$, so lines in this diagram are constant mass loss timescales. Now the threshold is much clearer and applies to all planets up to all planets with H/He envelopes. This also removes any effects from the somewhat arbitrary 15 $M_{\mathrm{\oplus}}$ cut.The sparsity of planets at low flux and high density is almost certainly a selection effect, since these are likely to be planets with long periods and small radii. However, the interesting result is that there is appears to be a critical mass loss timescale above which we do not find any  planets with H/He envelopes. In particular, all five of the inner Kepler-11 planets lie nicely along this threshold. Moreover, of the three planets that lie above the critical mass loss timescale, two are likely rocky. 

\begin{equation}\label{tmleq}
t_{\mathrm{loss}} = \frac{M_{\mathrm{p}}}{\dot{M}} = \frac{G M_{\mathrm{p}}^2}{\pi \epsilon R_{\mathrm{p}}^3 F_{\mathrm{XUV,E100}}} \frac{F_{\mathrm{\oplus}}}{F_{\mathrm{p}}}
\end{equation}

The dashed black line in Figure \ref{improvedfig} shows our best fit for this critical mass loss timescale. Equation \ref{tmleq} defines this mass loss timescale. Here $\epsilon=0.1$ is the mass loss efficiency, $F_{\mathrm{XUV,E100}} = 504$ $\mathrm{erg \, s^{-1} \, cm^{-2}}$ is the XUV flux at the Earth when it was 100 Myr old, and $F_{\mathrm{p}}$ is the current incident bolometric flux at a planet. We find a best fit with $t_{\mathrm{loss,crit}}\approx 12$ Gyr. However, while equation \ref{tmleq} accounts for the higher XUV fluxes at earlier times, it does not include the effects of larger radii at formation. The will reduce $t_{\mathrm{loss}}$ by at least another order of magnitude.

A similar mass loss threshold was proposed by \ct{Lecavelier2007}. Unfortunately, at that time there were relatively few transiting planets and no known transiting super-Earths. As a result, the authors we mostly limited to hot Jupiters from radial velocity surveys and were forced to use a scaling law to estimate radii. Here we are able to confirm the existence of a mass loss threshold and extend it all the way down to $\sim 2$ $M_{\mathrm{\oplus}}$.

This mass loss threshold could also help explain features in occurrence rate of planets found by \emph{Kepler}. \ct{Howard2011a} found that the frequency of 2-4 $R_{\mathrm{\oplus}}$ \emph{Kepler} planet candidates dropped off exponentially for periods within 7 days. This 7 day cutoff corresponds to an incident bolometric flux of 200 $F_{\mathrm{\oplus}}$. There are five planets with measured densities in figure \ref{shorelinefig} that lie above  200 $F_{\mathrm{\oplus}}$. Of these five, three planets are consistent with being rocky and two with being water-worlds; none of the five requires a H/He atmosphere to match its observed mass and radius. If all low mass planets orbiting within 7 days lose their H/He atmospheres, then their radii will shrink from 2-4 $R_{\mathrm{\oplus}}$ to $<$2 $R_{\mathrm{\oplus}}$. This could naturally explain the drop off in 2-4 $R_{\mathrm{\oplus}}$ candidates at short periods.

\subsection{Reproducing the Mass Loss Threshold}
\label{reproducesec}

\begin{figure*}[h!] 
  \begin{center}
    \includegraphics[width=6.in,height=4.3in]{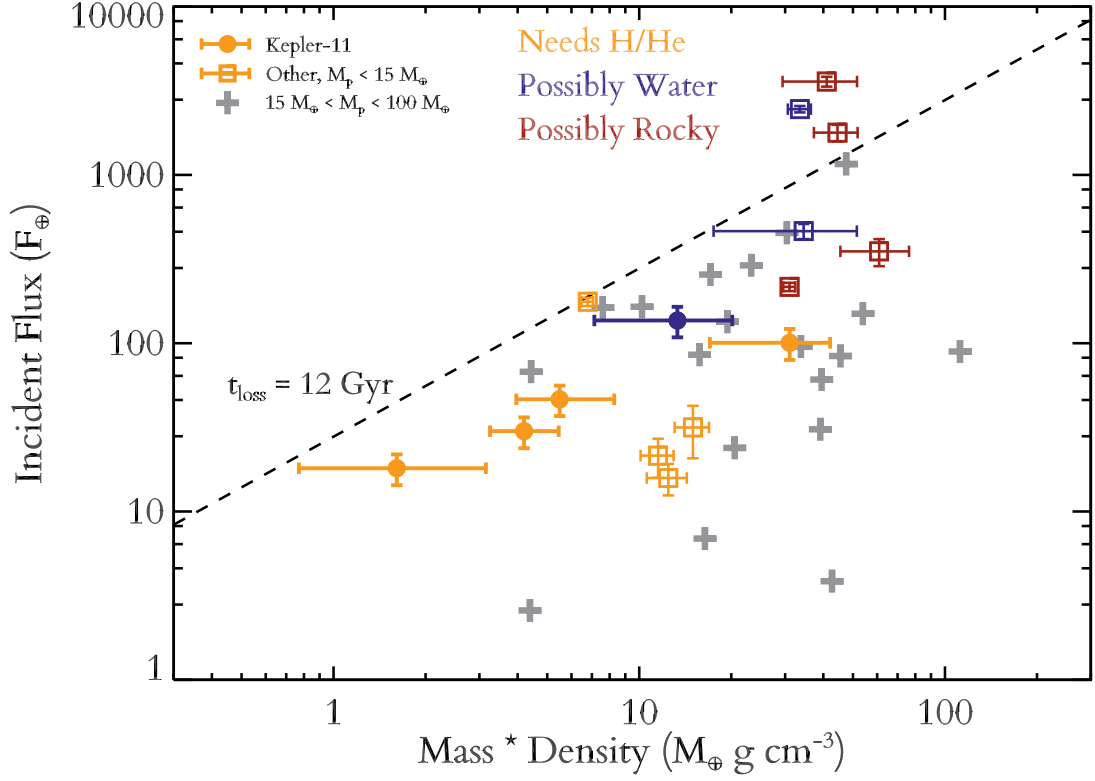}
  \end{center}
  \caption{Similar to Figure \ref{shorelinefig} except here we have multiplied the x-axis by planet mass. Once again, Kepler-11 planets are shown by filled circles, while open squares show the other extrasolar planets included in Figure \ref{mrfig}. Colors indicate possible compositions. Low mass planets that could be rocky are red, those that could be water-worlds are blue, and those that must have H/He are orange. For comparison, the gray crosses show all other transiting planets with measured masses greater than 15 $M_{\mathrm{\oplus}}$ and less than 100 $M_{\mathrm{\oplus}}$. There is a threshold in this diagram above which there are no observed transiting planets. Moreover, this threshold corresponds to a critical mass-loss timescale (see eq. \ref{fdeq}), as shown by the dashed black line. We discuss this threshold in the context of XUV driven mass loss in section \ref{fdsec}. In section \ref{reproducesec} and figure \ref{predictfig} we reproduce this threshold using our coupled thermal evolution and mass loss models. Finally in section \ref{constraintsec} and table \ref{masslimtab} we use this threshold to provide minimum masses for the broader population of \emph{Kepler} candidates without measured densities. \label{improvedfig}}
\end{figure*}

In order to fully examine whether the mass loss threshold in Figure \ref{shorelinefig} can be explained by atmospheric mass loss, we performed a small parameter study with $\sim$800 mass loss models across a wide range of initial masses, compositions, and incident fluxes. For each model we ran thermal evolution and mass loss starting at 10 Myr around a Sun-like star. We ran models with initial masses of 2, 4, 8, 16, 32, and 64 $M_{\mathrm{\oplus}}$. We assumed water-poor super-Earth compositions, meaning H/He envelopes on top Earth-like cores, with initial compositions of 1, 2, 5, 10, 20, and 40\% H/He. Finally we varied the incident bolometric flux from 10 to 1000 $F_{\mathrm{\oplus}}$, in order to cover the range of observed planets in Figures \ref{shorelinefig} and \ref{improvedfig}. We then recorded the resulting masses, densities, and compositions at various ages.

\begin{figure*}[h!] 
  \begin{center}
    \includegraphics[width=6.in,height=3.4in]{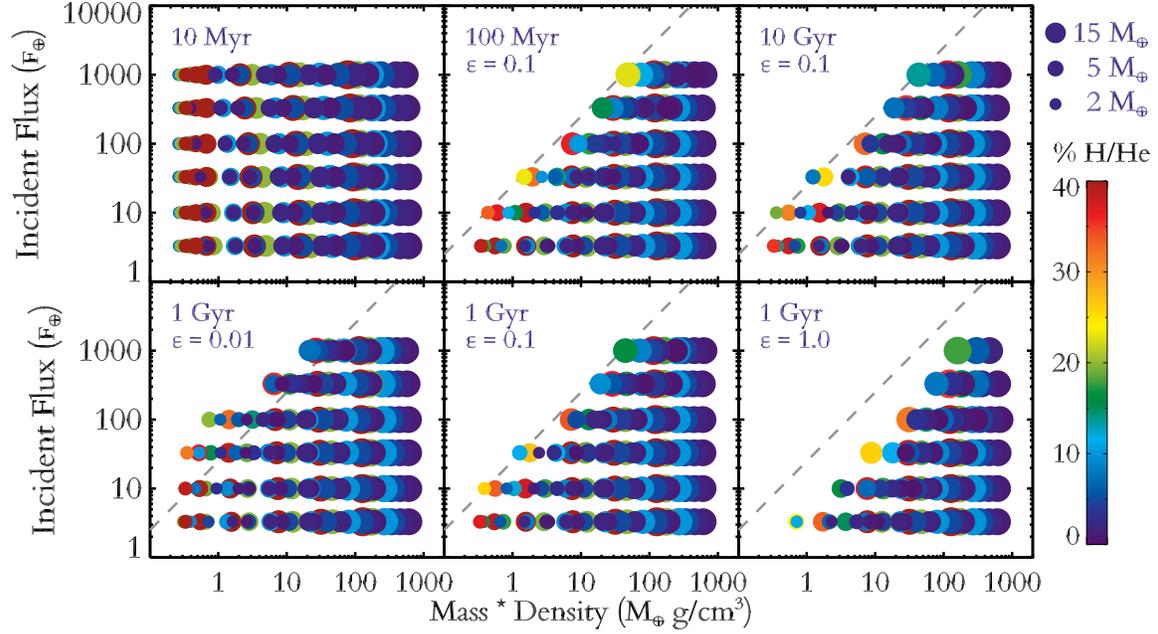}
  \end{center}
  \caption{This shows the results of $\sim$ ~1000 thermal evolution and mass loss model runs which reproduce the mass loss threshold seen in Figure \ref{improvedfig}. Each panel plots incident bolometric flux in $F_{\mathrm{\oplus}}$ vs. planet density $\times$ mass for different ages and mass loss histories. Each point is sized according to its mass and colored according to its composition, assuming a H/He atmosphere atop an Earth-like core. The top left panel shows the initial distribution of the models before any mass loss has taken place. The other two top panels show results at 100 Myr and 10 Gyr for our standard mass loss efficiency ($\epsilon=0.1$). Meanwhile the bottom panels show the results at 1 Gyr for three different mass loss efficiencies ranging from extremely inefficient ($\epsilon=0.01$), to extremely efficient mass loss ($\epsilon=1$). As planets lose mass, the points shrink, move to the right, and become bluer. The dashed line in panels 2-6, is the same as the black dashed line in Figure \ref{improvedfig} corresponding to critical mass loss timescale. The threshold in Figure \ref{shorelinefig} is well reproduced by models with $\epsilon=0.1$, which is also the approximate value implied by detailed models as discussed in Section \ref{masslosssec}. \label{predictfig}}
\end{figure*}

The results are shown in Figure \ref{predictfig}. As in Figure \ref{improvedfig}, each panel plots the total incident flux at the top of the atmosphere vs. the planet mass times density assuming different mass loss histories for our full suite of models. The size of each point indicates the mass of the planet, while the color indicates the fraction of its mass in the H/He envelope. The top left panel shows the initial distribution at 10 Myr before we start any mass loss. The other two top panels show the results at 100 Myr and 10 Gyr for our standard mass loss efficiency $\epsilon=0.1$. Meanwhile, the bottom panels show the results at 1 Gyr for a range of different efficiencies. These range from highly inefficient mass loss $\epsilon=0.01$,  to our standard efficiency $\epsilon = 0.1$, and finally extremely efficient mass loss $\epsilon=1$. In each panel, as planets cool and lose mass the points move to the right, shrink, and become bluer (less H/He). For reference, we have re-plotted our critical mass loss timescale from Figure \ref{improvedfig} in each of the result panels.

As we can see, models with mass loss do in general result in a threshold roughly corresponding to a critical mass loss timescale. Moreover, the mass loss threshold observed in Figure \ref{shorelinefig} is well reproduced by mass loss models with $\epsilon \approx 0.1$. This is similar to the efficiencies found by detailed models of mass loss from hot Jupiters in the energy-limited regime \cp{Murray-Clay2009}. This suggests that our assumption of comparable mass loss efficiencies for LMLD planets is reasonable. It is also apparent that the threshold already in place by 100 Myr, and  subsequent evolution has a relatively minor effect. We also examined the effect of beginning our parameter study at 100 rather than 10 Myr; however, this did not significantly affect the location of the threshold.

Previous mass loss evolution models \cp[e.g.,][]{Hubbard2007a, Hubbard2007b, Jackson2012, Owen2012} have also predicted mass loss thresholds. However, our models are the first to fully include the effects of coupled mass loss and thermal evolution for LMLD planets. We are able confirm and explain the observed threshold seen in Figure \ref{improvedfig} in a region of parameter space where most of the {\it Kepler} planets are being found.

\subsection{Constraints On Mass and Radius for the General Population}
\label{constraintsec}

If we use the critical mass loss timescale curve from Figure \ref{improvedfig} as a approximation for the observed mass loss threshold, then we can write down a simple expression for the threshold. This is shown in equation (\ref{fdeq}), which is valid for planets around Sun-like stars with  $F_{\mathrm{p}}< 500$ $F_{\mathrm{\oplus}}$. The 500 $F_{\mathrm{\oplus}}$ cut excludes highly irradiated rocky planets like Kepler-10b and CoRoT-7b. These planets may have once had volatiles in the past, but they are likely rocky today and so H/He mass loss is no longer relevant. This cut also excludes the region where energy-limited escape breaks down and mass loss becomes radiation and recombination limited \cp{Murray-Clay2009}

\begin{equation}\label{fdeq}
\rho M_{\mathrm{p}} \ge \frac{ 3\epsilon F_{\mathrm{XUV,E100}} } { 4 G } \frac{ F_{\mathrm{p}} } { F_{\mathrm{\oplus}} } t_{\mathrm{loss,crit}}
\end{equation}

The exciting implication of equation (\ref{fdeq}) is that we can use it to obtain lower limits on mass for the much larger population of \textit{Kepler} super-Earths and sub-Neptunes for which we do not have measured densities. This will help identify promising targets for follow-up work with radial velocity observations. This is shown in equation (\ref{masslimeq}). 

\begin{equation}\label{masslimeq}
M_{\mathrm{p}} \ge \sqrt{ \frac{\pi\epsilon F_{\mathrm{XUV,E100}}}{G} \frac{F_{\mathrm{p}}}{F_{\mathrm{\oplus}}} t_{\mathrm{loss,crit}}} \, R_{\mathrm{p}}^{3/2}
\end{equation}

Table \ref{masslimtab} applies equation (\ref{masslimeq}) to a list of \textit{Kepler} candidates smaller than 4 $R_{\mathrm{\oplus}}$ that are well suited to radial-velocity follow-up. We excluded any planets with $F_{\mathrm{p}}> 500$ $F_{\mathrm{\oplus}}$, since equation (\ref{masslimeq}) is not valid in that regime. Also, we limited the sample to only those planets with minimum radial velocity semi-amplitudes $K_{\mathrm{min}}>1.0 \: \mathrm{m \, s^{-1}}$ around stars with Kepler magnitude brighter than 13, since these will be the most promising for RV follow-up. In the end, this leaves us with a list of 38 likely detectable targets, eight of which (KOIs 104.01, 107.01, 123.01, 246.01, 262.02, 288.01, 984.01, and 1241.02) have $K_{\mathrm{min}}>2.0 \: \mathrm{m \, s^{-1}}$.

Finally, we can also use the mass loss threshold to find an upper limit on the radii of non-transiting planets from radial velocity surveys with $F_{\mathrm{p}} < 500$ $F_{\mathrm{\oplus}}$. This is done in equation (\ref{radiuslimeq}).

\begin{equation}\label{radiuslimeq}
R_{\mathrm{p}} \le ( \frac{G}{\pi\epsilon F_{\mathrm{XUV,E100}} t_{\mathrm{loss,crit}}}\frac{F_{\mathrm{\oplus}}}{F_{\mathrm{p}}} )^{1/3} M_{\mathrm{p}}^{2/3}
\end{equation}

\section{Discussion}
\subsection{Kepler-11:  Comparison to Formation Models, Implications for Migration}
\label{formationsec}

By itself, the constraints from mass loss do not tell us whether Kepler-11 is a system of water-poor super-Earths or water-rich sub-Neptunes. Instead we need to compare our estimates of the initial compositions to models of planet formation. By doing so we can examine whether our estimates of the original compositions for a water-poor super-Earth scenario are consistent with the maximum H/He fraction that can be accreted during in situ formation.

\ct{Ikoma2012} examine the accretion of H/He atmospheres onto the rocky cores of hot water-poor super-Earths. In particular, they examine the in situ formation of the Kepler-11 system. In addition to a planet's core mass and temperature, the amount of H/He accreted will depend strongly on the lifetime and dust grain opacity of the accretion disk. As with thermal evolution, the need to cool the rocky core can slow the contraction of the accreting atmosphere and limit the final H/He fraction. They are able to set hard upper limits on the initial compositions for in situ formation by assuming a grain-free, long-lived ($\sim 1$ Myr in the inner 0.2 AU) accretion disk and ignoring the delay in accretion due to cooling the core.

In particular, \ct{Ikoma2012} find that Kepler-11b could not have accreted more than 10\% of its mass in H/He if it formed in situ. Moreover, using a more typical disk lifetime of $10^5$ yr \cp{Gorti2009} and including the effect of cooling the core implies that Kepler-11b was $<1\%$ H/He at formation. On the other hand, in Table \ref{masslosstab} we showed that thermal evolution and mass loss models predict that if Kepler-11b is a water-poor super-Earth then it was $87\pm^{7}_{85}\%$ and at least $82\%$ at 10 Myr. Combined with the results of \ct{Ikoma2012}, this disfavors in situ formation of Kepler-11b. This result appears robust to any uncertainties in thermal evolution or mass loss models. Even if we only look after the period of run-away mass loss, at 3 Gyr Kepler-11b was still 10$\%$ H/He, the maximum allowed by \ct{Ikoma2012}. Likewise, we find that Kepler-11f was at least 10$\%$ H/He at 10 Myr, even though the \ct{Ikoma2012} models predict that it cannot have accreted its current composition of 4$\%$ H/He if it formed in situ. Furthermore, the co-planar, tightly packed, circular orbits in the system strongly suggest that it could have undergone type 1 migration \cp{Ida2010}. As a result, we disfavor in situ formation of the system.

If the Kepler-11 system did not form at its current location, then one possibility is that it formed at or beyond the snow-line and then Type 1 migrated to it is current location \cp{Rogers2011}. If this is the case, then it is likely a system of water-rich sub-Neptunes and water-worlds as discussed in Section \ref{watersec}. As we showed in Section \ref{watersec}, Kepler-11b is very stable to mass-loss if it is a water-world. If it was initially a water-rich sub-Neptune, it could have easily lost its H/He layer in the first few 100 Myr. Likewise, Kepler-11c-f are all consistent having formed as water-rich sub-Neptunes with $\sim20\%$ of their mass in H/He. 

The other possibility is that Kepler-11 is a system of water-poor super-Earths that has nonetheless undergone significant migration. For a grain-free accretion disk that lasts $10^6$ yr at 550 K, the critical mass for run-away accretion drops to 5 $M_{\mathrm{\oplus}}$ \cp{Ikoma2012}. This implies that Kepler-11b could possibly have formed as a water-poor super-Earth at or beyond the current orbit of 11f. Nonetheless, this assumes a completely grain-free long-lived disk, which may not be realistic. Furthermore, this scenario still requires that Kepler-11b was $\sim$90\% H/He when it formed, while all the other planets in the system are consistent with more modest initial compositions. As a result, we favor the water-rich sub-Neptune scenario.

\subsection{Kepler 11: Mass Loss and Orbital Stability}
\label{hillsec}

\begin{figure*}[h!] 
  \begin{center}
    \includegraphics[width=6.in,height=4.3in]{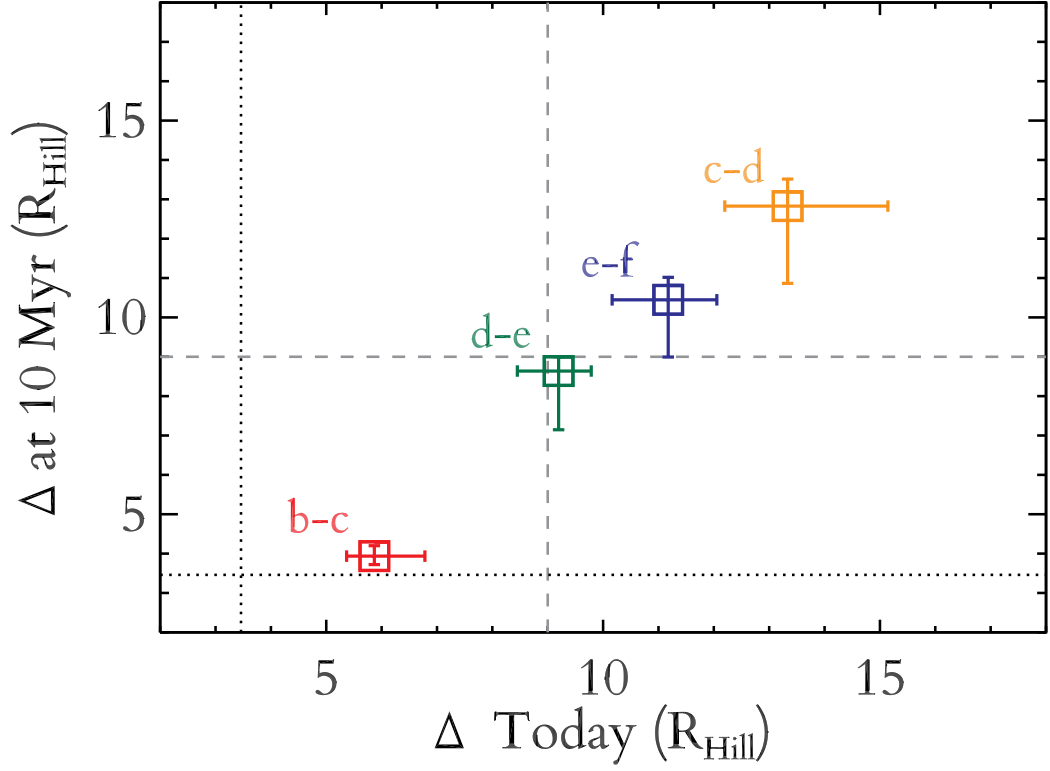}
  \end{center}
  \caption{Separation between adjacent pairs of planets in Kepler-11, in terms of their mutual Hill spheres $\Delta$. The x-axis shows the separations in terms of the current Hill spheres, while the y-axis shows the predicted Hill spheres when the system was 10 Myr old assuming a water-poor super-Earth scenario and that the planets remained stationary. The dashed lines show the approximate $\Delta>9$ stability threshold for five planet systems from \ct{Smith2009}. Likewise, the dotted lines show the $\Delta>2\sqrt{3}$ stability threshold for two planet systems from \ct{Gladman1993}. Although the d-e pair does drop below the $\Delta>9$ threshold at 10 Myr, both the c-d and e-f pairs remain above it potentially stabilizing the system. More importantly, the b-c pair drops dangerously close to the $\Delta>2\sqrt{3}$ critical threshold for dynamical stability in two-planet systems. This is another reason we disfavor a water-poor super-Earth scenario for Kepler-11b.\label{hillfig}}
\end{figure*}

One possible result of significant mass loss is that it could impact the orbital stability of closely packed multi-planet systems like Kepler-11. Although this system is stable in its current configuration, it might not be with the initial masses determined by our models. One relatively simple stability check is to calculate the separation between pairs of planets in terms of their mutual Hill spheres. Figure \ref{hillfig} plots the separation in mutual Hill spheres ($\Delta$) between adjacent pairs of planets at both 10 Myr and the present, assuming a water-poor super-Earth composition.

\ct{Smith2009} found that systems with five or more planets tended to de-stabilize when $\Delta<9$. This threshold is shown as dashed gray lines in figure \ref{hillfig}. Although Kepler-11b-c currently lies well below this threshold, \ct{Lissauer2011a} showed that the system is nevertheless stable today because planets b and c are dynamically decoupled from the other four planets and so act more like a two planet system. For two planet systems the absolute minimum stable separation is $\Delta=2\sqrt{3}=3.46$ \cp{Gladman1993}. This second stability threshold is shown by the dotted lines in figure \ref{hillfig}. The Hill radius goes as $M_p^{1/3}$, as a result the change in $\Delta$ from mass loss is relatively modest; nonetheless, the stability of the system is in danger. At 10 Myr, Planets d-e do lie below the approximate $\Delta>9$ stability threshold; however, both pairs on either side of d-e are still relatively stable which may help stabilize the system. More importantly, the separation of planets b-c at 10 Myr skirts dangerously close, $\Delta=3.8\pm^{0.5}_{0.4}$, to the critical $\Delta>2\sqrt{3}$ stability threshold. More detailed modeling needs to be done to assess the impact of mass loss on orbital stability; nonetheless, the $\Delta>2\sqrt{3}$ stability threshold provides another strong reason to be skeptical of a water-poor super-Earth scenario for Kepler-11b.

The major caveat to this stability analysis is that we assume that all of the orbits are stationary even as the planets lose mass. This is motivated by \ct{Adams2011}, which showed that in the presence of a modest planetary magnetic field XUV driven mass loss from hot Jupiters tends to come out along the magnetic poles. Assuming that the magnetic field is sufficiently strong, dipolar, and perpendicular to the plane of the orbit, then mass loss won't have any impact on the orbit. In general however, the directionality of mass loss will be an extremely complicated problem determined by the interaction of the ionized hydrodynamic wind, the planetary magnetic field, and the stellar wind. \ct{Boue2012} showed that if the mass loss is directed in the plane of the orbit, then it can have a significant impact on both semi-major axis and eccentricity.

\subsection{Future Work}
Further confirmation of the mass loss threshold will depend on getting reliable mass estimates for more LMLD planets. Fortunately, there is a large population of super-Earth sized planets in multi-planet systems found by \textit{Kepler} \cp{Lissauer2011b}. For some of these systems TTV can be used to determine masses \cp{Agol2005}. Moreover, this will become possible for more systems as more quarters of data are collected. Likewise, as more quarters of transit data are analyzed, previous mass constraints from TTV will become more precise. For Kepler-11 this will allow tighter constraints on both the current and past compositions.

In order to better understand mass loss, there is also a strong need to acquire more XUV observations of young ($\sim10$ Myr) stars. Currently the best estimates of EUV fluxes are for G and K stars older than 100 \cp{Ribas2005,Sanz-Forcada2010}. Meanwhile, planet formation ends and mass loss becomes important after a few Myr \cp{Calvet2002, Alexander2006}. A large amount of planetary mass will be lost in the first 100 Myr and this will depend strongly on the stellar XUV flux. \ct{Jackson2012} recently found that x-ray fluxes saturate for stars younger than 100 Myr; however, this needs to be investigated at other wavelengths. Likewise, more UV observations of transiting exospheres are needed. Currently, we only have observations for a handful of hot Jupiters \cp{Vidal-Madjar2004, Lecavelier2010, Lecavelier2012}. More observations are needed, especially for Neptune and super-Earth sized planets.

Likewise, it is also important get more observations of XUV fluxes and flares from M dwarfs, which are known to be highly active \cp{Reiners2012}. This is particularly important for super-Earths in the habitable zones of late M dwarfs where the stability of habitable atmospheres could depend on XUV driven mass loss. The three planets orbiting M-stars in figure \ref{improvedfig} all lie an order of magnitude below the threshold for Sun-like stars. This could be due to mass loss from X-ray flares, but without a larger sample size it is impossible to say. The Recently \ct{France2012} obtained the first FUV spectrum for GJ876; however, more data are needed. Likewise, non-equilibrium mass loss models need to be developed to understand the impact of flares.

In the area of modeling mass loss, we have utilized a relatively simple model that appears to work well for hot Jupiters, although mass loss rates have only been constrained for HD 209458b \cp{Vidal-Madjar2004} and HD 189733b \cp{Lecavelier2010, Lecavelier2012}.  On a broader scale, there may be evidence for a lost population of hot Jupiter planets at very high levels of XUV irradiation \cp[e.g.,][]{Davis2009}.  Since we have suggested that H/He mass loss may be more important for more modestly irradiated LMLD planets than for average hot Jupiters (since the smaller planets have much smaller H/He masses, mass loss for LMLD planets can actually \emph{transform} the very structure of the planets), we encourage further detailed models to quantitatively access mass loss from these atmospheres.

\section{Conclusions}

In order to better understand the structure, history, and formation of low-mass planets, we constructed coupled thermal evolution and mass loss models of water-poor super-Earths, water-worlds, and water-rich sub-Neptunes. The Kepler-11 system represents a new class of low-mass low-density planets that offers a unique test-bed for such models and gives us powerful insights on planet formation and evolution. Applying this understanding more broadly, we find a relation between a planet's mass, density, and its incident flux that matches the observed population. Moreover, this threshold can help constrain the properties of hundreds of planets.  Our primary conclusions are:

\begin{itemize}
\item XUV-driven hydrogen mass loss coupled with planetary thermal evolution is a powerful tool in understanding the composition and formation of low-mass low-density planets.

\item A coupled model is essential for this work, due to the much larger planetary radii in the past, when XUV fluxes were significantly higher.

\item In situ formation of the Kepler-11 system is disfavored, instead it could be a system of water-rich sub-Neptunes that formed beyond the snow line.

\item If Kepler-11b is a water-poor super-Earth then it likely formed with $\sim 90\%$ H/He beyond 0.25 AU. We believe this is unlikely and instead show that Kepler-11 b-f all could have originated as water-rich sub-Neptunes with $\sim 20\%$ H/He initially. If this is the case, Kepler-11b could have lost its H/He envelope and become a water-world today for a wide range of initial masses and compositions. 

\item There is a sharp observed threshold in incident flux vs. planet density times mass above which we do not find planets with H/He envelopes.  To date, low-density planets have not been found above this threshold.

\item This mass loss threshold is well reproduced by our coupled thermal evolution and mass loss models.


\item This threshold can be used to provide limits on planet mass or radius for the large population of low-mass low-density planets without measured densities.
\item In particular, we have identified promising \emph{Kepler} targets for RV follow-up.
\end{itemize}

\acknowledgements{We would like to thank Nadine Nettelmann for help implementing the water EOS; also Masahiro Ikoma and Yasunori Hori for providing us their constraints from formation. EDL would like to thank Dan Fabrycky, Kevin Zahnle, Doug Lin, Eugene Chiang, James Owen, and Eliza Kempton for many helpful conversations. This research has made use of the Exoplanet Orbit Database and the Exoplanet Data Explorer at exoplanets.org." We acknowledge the support of NASA grant NNX09AC22G.} 

\bibliographystyle{apj}
\bibliography{myreferences}

\LongTables
\begin{deluxetable*}{ccccccccc}[h!]
  \footnotesize
  \tablecaption{Minimum Masses and Densities for Selected KOI Candidates}
  \tablewidth{0pt}
  \tablehead{
   \colhead{KOI} & \colhead{KepMag} & \colhead{P (day)} & \colhead{$F_{\mathrm{p}}$ ($F_{\mathrm{\oplus}}$)} & \colhead{$R_{\mathrm{p}}$ ($R_{\mathrm{\oplus}}$)}  & \colhead{$\rho_{\mathrm{min}}$ (g cm$^{-3}$) } & \colhead{$M_{\mathrm{p,min}}$ ($M_{\mathrm{\oplus}}$)} & \colhead{$K_{\mathrm{min}}$ (m s$^{-1}$)} 
}
  \startdata

$70.01$ & $12.50$ & $10.85$ & $80.3$ & $3.09$ & $0.73$ & $3.9$ & $1.21$
\\\\
$70.02$ & $12.50$ & $3.69$ & $343.5$ & $1.92$ & $3.0$ & $3.9$ & $1.75$
\\\\
$85.01$ & $11.02$ & $5.85$ & $403.4$ & $2.35$ & $2.4$ & $5.8$ & $1.83$
\\\\
$94.02$ & $12.21$ & $10.42$ & $209.9$ & $3.43$ & $1.0$ & $7.4$ & $1.92$
\\\\
$104.01$ & $12.90$ & $2.50$ & $233.4$ & $3.36$ & $1.1$ & $7.5$ & $4.08$
\\\\
$105.01$ & $12.87$ & $8.98$ & $130.3$ & $3.35$ & $0.82$ & $5.6$ & $1.94$
\\\\
$107.01$ & $12.70$ & $7.25$ & $301.7$ & $3.09$ & $1.4$ & $7.6$ & $2.29$
\\\\
$110.01$ & $12.66$ & $9.94$ & $220.5$ & $2.92$ & $1.3$ & $5.9$ & $1.62$
\\\\
$115.02$ & $12.79$ & $7.12$ & $409.1$ & $1.88$ & $3.4$ & $4.2$ & $1.29$
\\\\
$117.02$ & $12.49$ & $4.90$ & $436.9$ & $1.70$ & $4.1$ & $3.7$ & $1.25$
\\\\
$122.01$ & $12.35$ & $11.52$ & $108.8$ & $2.78$ & $1.0$ & $3.9$ & $1.06$
\\\\
$123.01$ & $12.37$ & $6.48$ & $461.4$ & $2.64$ & $2.2$ & $7.4$ & $2.46$
\\\\
$124.01$ & $12.94$ & $12.69$ & $227.8$ & $3.00$ & $1.2$ & $6.3$ & $1.65$
\\\\
$246.01$ & $10.00$ & $5.39$ & $404.8$ & $2.53$ & $2.2$ & $6.5$ & $2.27$
\\\\
$257.01$ & $10.87$ & $6.88$ & $308.6$ & $2.61$ & $1.8$ & $5.9$ & $1.80$
\\\\
$262.02$ & $10.42$ & $9.37$ & $491.8$ & $2.79$ & $2.1$ & $8.3$ & $2.20$
\\\\
$277.01$ & $11.87$ & $16.23$ & $177.4$ & $3.82$ & $0.79$ & $8.0$ & $1.91$
\\\\
$280.01$ & $11.07$ & $11.87$ & $154.9$ & $2.52$ & $1.3$ & $4.0$ & $1.16$
\\\\
$281.01$ & $11.95$ & $19.55$ & $192.3$ & $3.46$ & $0.95$ & $7.2$ & $1.99$
\\\\
$285.01$ & $11.57$ & $13.74$ & $180.5$ & $3.38$ & $0.96$ & $6.7$ & $1.61$
\\\\
$288.01$ & $11.02$ & $10.27$ & $433.9$ & $3.11$ & $1.6$ & $9.2$ & $2.14$
\\\\
$291.02$ & $12.85$ & $8.12$ & $247.8$ & $2.14$ & $2.2$ & $3.9$ & $1.25$
\\\\
$295.01$ & $12.32$ & $5.31$ & $339.7$ & $1.77$ & $3.4$ & $3.5$ & $1.22$
\\\\
$297.01$ & $12.18$ & $5.65$ & $482.1$ & $1.65$ & $4.6$ & $3.7$ & $1.25$
\\\\
$301.01$ & $12.73$ & $6.00$ & $399.2$ & $1.75$ & $3.8$ & $3.7$ & $1.17$
\\\\
$323.01$ & $12.47$ & $5.83$ & $166.2$ & $2.17$ & $1.7$ & $3.3$ & $1.21$
\\\\
$984.01$ & $11.63$ & $4.28$ & $259.7$ & $3.19$ & $1.2$ & $7.4$ & $2.91$
\\\\
$987.01$ & $12.55$ & $3.17$ & $404.8$ & $1.28$ & $6.1$ & $2.3$ & $1.03$
\\\\
$1117.01$ & $12.81$ & $11.08$ & $327.5$ & $2.20$ & $2.4$ & $4.7$ & $1.13$
\\\\
$1220.01$ & $12.99$ & $6.40$ & $441.4$ & $1.95$ & $3.4$ & $4.6$ & $1.52$
\\\\
$1241.02$ & $12.44$ & $10.50$ & $485.3$ & $3.84$ & $1.3$ & $13.3$ & $3.17$
\\\\
$1597.01$ & $12.68$ & $7.79$ & $423.6$ & $2.67$ & $2.0$ & $7.2$ & $1.86$
\\\\
$1692.01$ & $12.56$ & $5.96$ & $175.9$ & $2.65$ & $1.3$ & $4.6$ & $1.61$
\\\\
$1781.01$ & $12.23$ & $7.83$ & $61.3$ & $3.29$ & $0.58$ & $3.7$ & $1.38$
\\\\
$1781.02$ & $12.23$ & $3.00$ & $219.6$ & $1.94$ & $2.4$ & $3.2$ & $1.63$
\\\\
$1921.01$ & $12.82$ & $16.00$ & $172.9$ & $3.09$ & $1.0$ & $5.7$ & $1.28$
\\\\
$1929.01$ & $12.73$ & $9.69$ & $251.7$ & $2.00$ & $2.4$ & $3.6$ & $1.11$
\\\\
$2067.01$ & $12.58$ & $13.24$ & $347.2$ & $2.97$ & $1.6$ & $7.6$ & $1.69$
\\\\

  \enddata
  \label{masslimtab}
  \tablecomments{Minimum masses and densities given by equation (\ref{masslimeq}) for KOI candidates selected for being promising targets for RV follow-up. In order to focus on promising candidates we cut the sample to planets with $K_{\mathrm{min}}>1.0 \: \mathrm{m \, s^{-1}}$ around stars with KepMag $<$ 13. We only included planets with $F_{\mathrm{p}}< 500$ $F_{\mathrm{\oplus}}$ and $R_{\mathrm{p}}< 4$ $R_{\mathrm{\oplus}}$. This leaves us with 38 high-quality targets, eight of which have $K_{\mathrm{min}}>2.0 \: \mathrm{m \, s^{-1}}$.}
\end{deluxetable*}

\end{document}